\documentstyle[12pt,setspace,subeqn,subeqnarray]{article}
\catcode `@=11 \@addtoreset{equation}{section} \catcode `@=12
\begin{document}
\def\be{\begin{equation}}
\def\ee{\end{equation}}
\def\ba{\begin{array}{l}}
\def\ea{\end{array}}
\def\beq{\begin{eqnarray}}
\def\eeq{\end{eqnarray}}
\def\eq#1{(\ref{#1})}
\def\del{\partial}
\def\Del{\nabla}
\def\A{{\cal A}}

\renewcommand\arraystretch{1.5}

\begin{flushright}
TIFR-TH-00/30\\
hep-th/0006043
\end{flushright}

\begin{center}
\vspace{3 ex}
{\bf NONCRITICAL STRINGS, RG FLOWS AND HOLOGRAPHY} \\
\vspace{8 ex}
Avinash Dhar$^*$ and Spenta R. Wadia$^\dagger$ \\
{\sl Department of Theoretical Physics} \\
{\sl Tata Institute of Fundamental Research,} \\
{\sl Homi Bhabha Road, Mumbai 400 005, INDIA.} \\
\vspace{15 ex}
\pretolerance=1000000
\bf ABSTRACT\\
\end{center}
\vspace{2 ex}

We derive an RG flow equation that is satisfied by the regularized
partition function for noncritical strings in background fields. The
flow refers to change in the position of a ``boundary'' in the
liouville direction. The boundary is required to regularize the
ultraviolet divergences in the partition function coming from
integration over world-sheets of arbitrarily small area. From the
point of view of the target space effective gravitational action that
the partition function evaluates on-shell, the boundary regularizes
{\it infrared} divergences coming from the infinite volume of the
liouville direction. The RG flow equation that we obtain looks very
much like the Hamilton-Jacobi constraint equation that an on-shell
gauge-fixed gravitational action must satisfy.
    
\vfill
\hrule
\vspace{0.5 ex}
\leftline{$^*$ adhar@tifr.res.in}
\leftline{$^\dagger$ wadia@tifr.res.in}
\clearpage

\doublespace
\section{INTRODUCTION}

One of the central results that has emerged from the studies of the
AdS/CFT correspondence \cite{M,GKP,W} between $5$-dimensional gravity
in the AdS bulk and $4$-dimensional Yang-Mills theory on its boundary
is the identification of the renormalization scale of the latter with
the radial coordinate of the AdS bulk and the radial evolution of the
$5$-dimensional fields with RG flows of the couplings in the
$4$-dimensional Yang-Mills theory
\cite{A,AG,BKL,BKLT,GPPZ,DZ,KPW,KLM,KW,FGPW,GPPZa,ST,DeFGK,BVV,PS,CS}. 
A conjectured generalization of this correspondence to
$4$-dimensional boundary theories which include gravity \cite{RS,V} has 
also recently been studied in the context of Randall-Sundram type
brane-world scenarios \cite{G,S,GRS,CEH,DGP} and the cosmological constant
problem \cite{HDKS,KSS,VV,Gub}.

The existence of a connection between RG flows in a D-dimensional
theory, which includes gravity, and gravitational equations in
(D+1)-dimensions, was recognized and pointed out quite sometime back
in the perturbative studies of noncritical string theory \cite{DDW}.
Perturbative noncritical string theory is formulated as D-dimensional
matter coupled to 2-d quantum gravity \cite{P}. As is well-known, in
this formulation of string theory the extra coordinate of the
(D+1)-dimensional space is related to the conformal degree of freedom
of the world-sheet metric \cite{DNW,BL,DJNW} and world-sheet
gravitational dressing \cite{KPZ} of the various $\sigma$-model
couplings gives rise to their dependence on this extra coordinate. A
connection between the RG flows of the D-dimensional fields and
gravitational equations in (D+1)-dimensions arises \cite{DDW} because
the dependence of the $\sigma$-model couplings on the conformal mode
of the world-sheet metric is determined by gravitational equations in
(D+1)-dimensions. In recent years this connection has been made
precise in the context of AdS/CFT correspondence. The purpose of this
note is to reexamine and further expand on the world-sheet approach of
noncritical string theory to holographic RG in the light of these
recent advances. The main advantage of this approach is that it
provides a natural setting for the discussion of a generic holographic
RG connection between D-dimensional boundary theories that contain
gravity and (D+1)-dimensional gravitational dynamics. The world-sheet
approach also provides a systematic handle on stringy (i.e. $\alpha'$)
as well as string loop corrections.

The key organizing principle in the first quantized approach to
noncritical string theory is the requirement of world-sheet
reparametrization invariance. In the background gauge-fixing method, a
prescription for integrating over the 2-d metric which ensures this
requirement of reparametrization invariance automatically ensures Weyl
invariance with respect to the 2-d fiducial metric. In this approach,
therefore, Weyl invariance with respect to the 2-d fiducial metric
emerges as the principal consistency requirement which is needed to
ensure world-sheet reparametrization invariance. For example, it is
this requirement that determines the gravitational dressing of the
$\sigma$-model couplings.

In this note we will consider the $\sigma$-model partition function of
noncritical strings propagating in background fields. This partition
function is in general not well-defined because of divergent
contributions to it arising from correlators of microscopic loop
operators whose liouville wavefunctions are not normalizable. Although
this is an ultraviolet (small area) divergence on the world-sheet,
from the (D+1)-dimensional target space point of view it is an {\it
infrared} (large volume) divergence. It can be regularized by
introducing a cut-off on the integration over the liouville zero mode.
This is very much like the infrared regulator needed in the radial
direction to evaluate the on-shell gravity action in AdS space. This
way of regularizing the partition function introduces a ``boundary''
in the liouville direction. The regularized partition function depends
on the location of the liouville boundary only implicitly through the
values of the dressed couplings at the boundary. We will show here
that a change in the location of the boundary gives rise to an RG flow
equation for the partition function which looks exactly like the
Hamilton-Jacobi constraint which an on-shell boundary gravitational
action is expected to satisfy \cite{Pol,BVV}.

The plan of this paper is as follows. In the next section we first
summarize the main results \cite{DDW} from the first quantized
approach to noncritical string theory as D-dimensional matter coupled
to 2-d gravity. We then discuss in detail the interpretation of the
dependence of the $\sigma$-model couplings on the extra coordinate as
RG dependence in the D-dimensional theory. In Sec. 3 we argue that the
$\sigma$-model couplings should more correctly be interpreted as
defining a boundary value problem in a (D+1)-dimensional gravity
theory. We explain how the boundary arises from the need to regularize
world-sheet ultraviolet divergences in the calculation of the
partition function. We then show that, consistent with its
interpretation as an on-shell boundary action, the regularized
partition function staisfies a flow equation which looks very much
like a Hamilton-Jacobi constraint equation which an on-shell boundary
gravitational action is expected to satisfy. We end in Sec. 4 with
some concluding remarks.
 
\section{NONCRITICAL STRINGS AND HOLOGRAPHIC RG}

In this section we first briefly summarize some old and rather
well-known results from noncritical string theory. We then discuss the
RG scale dependence interpretation of the gravitational dressing of
the couplings. For simplicity we restrict the discussion to bosonic
string, but extension to superstring is straightforward.

The starting point of the first quantized approach to noncritical
strings in background fields is the world-sheet reparametrization
invariant action
\beq
S &=& {1 \over 8 \pi \alpha'} \int d^2 \xi {\sqrt g} 
\bigg[ \del_\alpha X^\mu \del_\beta X^\nu  \left( g^{\alpha \beta} 
G_{\mu \nu}(X(\xi))+ \epsilon^{\alpha \beta} 
B_{\mu \nu}(X(\xi))\right) \nonumber \\ [3mm] 
&&\displaystyle\hbox{~~~~~~~~~~~~~~~~~~~~} + \alpha' R^{(2)} \Phi(X(\xi)) 
+ T(X(\xi)) + \cdots \bigg]. 
\label{twoone}
\eeq 
Here $x^\mu$'s, which are the zero modes of $X^\mu(\xi)$'s,
parametrize a D-dimensional space with metric $G_{\mu\nu}(x)$ and
other fields. The Polyakov path integral formally defines the
partition function
\beq
Z[G_{\mu \nu}, \Phi, B_{\mu \nu}, \cdots] = \int [{\cal D} g_{\alpha 
\beta}] [{\cal D} X^\mu] \ e^{-S} 
\label{threeone} 
\eeq
which is a functional of the D-dimensional couplings $G_{\mu \nu}$, 
$\Phi$,  $B_{\mu \nu}$, etc. 

\vspace{2 ex}
\noindent {\bf Gravitational Dressing}
\vspace{1 ex}

In the quantum theory the various $\sigma$-model couplings get dressed
by 2-d gravity. A reparametrization invariant prescription for
determining these gravitational dressings is the following. One first
fixes the conformal gauge
$$g_{\alpha\beta}=e^{\phi(\xi)} \hat g_{\alpha\beta}.$$ 
Here $\phi(\xi)$ is the liouville mode and $\hat g_{\alpha\beta}$ is a
fiducial metric that depends on the moduli of the Riemann surface over
which the action in (\ref{twoone}) is defined. One then makes a
transformation in the functional integral from the liouville mode
$\phi(\xi)$ to a field $\eta(\xi)$ with gaussian measure which, in the
absence of the background fields, has the following action \cite{D,DK}
\beq
{1 \over 8 \pi} \int d^2 \xi \sqrt{\hat g} 
\left( \hat g^{\alpha\beta} \del_\alpha \eta \del_\beta \eta  
+ Q \hat R^{(2)} \eta \right) 
\label{twotwo}
\eeq 
where $Q = \sqrt {(25-D)/3}$. When background fields are switched on, 
in the presence of 2-d gravity they get dressed, that is they become 
functions of $\eta$ \cite{DNW,DDW,DJNW}. Thus
$$G_{\mu\nu}(x) \rightarrow G_{\mu\nu}(x, \eta), \ \ \Phi(x)
\rightarrow \Phi(x, \eta), \ \ \cdots$$ 
The $\eta$-dependence of the various fields is fixed by demanding that
the above procedure preserve world-sheet reparametrization
invariance. In particular, this means that the final results should be
invariant under Weyl transformations of the fiducial metric $\hat
g_{\alpha\beta}$. This leads to the familiar beta-function equations
for the dressed fields
\beq 
0 &=& R_{MN} + 2 \Del_M \Del_N \Phi - {1 \over 4} H_{MPL} {H_N}^{PL} 
+ \cdots 
\label{twothreea} \\ [3mm] 
0 &=& \Del^P H_{PMN} - 2 \Del^P \Phi H_{PMN} 
+ \cdots
\label{twothreeb} \\ [3mm] 
0 &=& {D-25 \over 3 \alpha'} - R^{(D+1)} -  4 \Del^P \Del_P \Phi + 
4 \Del_P \Phi \Del^P \Phi \nonumber \\   
&& ~~~~~~~~~~~~~~~~~~~~~~~ + {1 \over 12} H_{MNP} H^{MNP} + \cdots 
\label{twothreec}
\eeq 
etc. Here the indices $M, N$ etc. run over $\mu, \eta$ and the dots
represent $\alpha'$ and string loop corrections. These equations have
(D+1)-dimensional general covariance. 

\vspace{2 ex}
\noindent {\bf RG Flows}
\vspace{1 ex}

Although we considered D-dimensional matter in the above discussion,
these considerations actually apply to any matter coupled to 2-d
gravity. A term in the action of the form
$$\sum_i \int d^2\xi \sqrt g \ \lambda^i O_i(X(\xi), 
g_{\alpha \beta}(\xi)),$$ 
where $O_i(X(\xi), g_{\alpha \beta}(\xi))$ is a local
operator constructed from the matter fields $\{X(\xi)\}$, gets dressed to
$$\sum_i \int d^2\xi \sqrt {\hat g} \ \lambda^i(\eta(\xi)) 
O_i(X(\xi), \hat g_{\alpha \beta}(\xi))$$ 
and the $\eta$-dependence of the dressed coupling, $\lambda^i(\eta)$,
is determined by an appropriate vanishing beta-function condition. In
case the couplings $\lambda^i$ correspond to a CFT coupled to 2-d
gravity, the $\lambda^i(\eta)$ are independent of $\eta$. This
identifies CFT's as special points in the space spanned by the set of
all possible couplings $\lambda^i$, the so-called theory space.  The
more general case in which the couplings get dressed can be interpreted 
as giving RG flows between these special points corresponding to CFT's.
 
It is important to emphasize here that the RG flow that we are talking
about is not due to changes of cut-off in the 2-d QFT of $X(\xi)$'s
and $\eta(\xi)$. Although this cut-off is needed to do computations,
the vanishing beta-function conditions ensure that the couplings do
not depend on it. The RG flows that we are talking about are similar
to finite size scaling. The size is here provided by the invariant
area of the world-sheet, or its conjugate, the world-sheet
cosmological constant \cite{D,DK,DDW}. In the conformal gauge, the
flows thus correspond to the response of the couplings to changes of
the physical scale brought about by shifts of the liouville mode, and
hence of $\eta$. The vanishing beta-function equations describe how
the dressed couplings change with precisely these shifts of $\eta$. In
general, shifts of $\eta$ do not produce any simple scalings of the
couplings, except near the points in theory space described by CFT's.
The trajectories in theory space given by the dressed couplings
describe RG flows between two such points. An example of such an RG
flow between two $c<1$ minimal models has been discussed in detail in
\cite{DDW} where an explicit kink solution is given which describes RG
flow due to a nearly marginal perturbation. \footnote{RG flows in
two-dimensional theories coupled to gravity have also been considered
in \cite{KKP,ST}.}

For D-dimensional matter coupled to 2-d gravity, constant shifts of 
$\eta$ describe RG flows in a D-dimensional effective theory of
gravity. To see this, consider the (D+1)-dimensional gravity theory,
obtained after dressing by 2-d gravity, in the gauge
\footnote{As we shall discuss in the next section, it is natural to 
guess that the partition function of noncritical strings in background 
fields evaluates a (D+1)-dimensional gravitational action on-shell in 
this gauge.}  
\beq
G_{\eta\eta}=1, \ \ G_{\eta\mu}=0. 
\label{twofour}
\eeq
In this gauge, the (D+1)-dimensional metric is given by
$$ds_{D+1}^2=d\eta^2 + ds_D^2,$$
where the metric in a constant $\eta$ D-dimensional slice is given by
$$ds_D^2=G_{\mu\nu}(x, \eta) dx^\mu dx^\nu.$$  
A shift in $\eta$ produces a change in the D-dimensional metric which
is dictated by the (D+1)-dimensional gravitational equations. In
general, such a change generates a {\it local} change of scale in the
D-dimensional world and hence corresponds to a {\it local}
generalization of the usual RG flows. In the particular case that
$$G_{\mu\nu}(x, \eta)=\Omega(\eta) G_{\mu\nu}(x)$$ 
and $\Omega$ is a monotonic function of $\eta$, shifts in $\eta$ give
global changes of scale in the D-dimensional world and hence in this
case we recover the standard RG flows in the D-dimensional theory.

\section{THE FLOW EQUATION}

In this section we will argue that the partition function for
noncritical strings evaluates a (D+1)-dimensional gravitational action
on-shell, in the gauge (\ref{twofour}), for solutions with a boundary.
Consistent with this interpretation, we will derive a flow equation
for the partition function, which looks just like the Hamilton-Jacobi
constraint that the on-shell gauge-fixed (D+1)-dimensional action must
satisfy for ensuring full (D+1)-dimensional general covariance.

The partition function is formally given by the functional integral in
(\ref{threeone}) where $S$ is the action in (\ref{twoone}). Now, it is
well-known that the equations in (\ref{twothreea})-(\ref{twothreec})
can be derived from a (D+1)-dimensional gravitational action. Since
these equations also determine the 2-d gravitational dressing of the
$\sigma$-model couplings, it would seem consistent to identify the
partition function in (\ref{threeone}) with the (D+1)-dimensional
gravitational action evaluated on-shell. There is, however, a problem
in this identification, which we will now discuss.

The set of equations (\ref{twothreea})-(\ref{twothreec}) are partial
differential equations of second or higher order, depending on the
order to which terms are retained in $\alpha'$ expansion. At the
lowest order in $\alpha'$, the equations are second order differential
equations in $\eta$ for the dressed couplings.  A general solution for
these equations depends on two independent functions of $x$ for each
of the background fields. For example, for the dilaton the general
solution depends on $\Phi_0(x)$ and $\Phi_0'(x)$, which may be taken
to be respectively the value of the dressed field $\Phi(x, \eta)$ at
some point $\eta=\eta_0$ and the value of its derivative $\del_\eta
\Phi(x, \eta)$ at $\eta_0$. Thus, at the lowest order in $\alpha'$,
the (D+1)-dimensional on-shell action should depend on two independent
functions of $x$ for each of the background fields. The partition
function in (\ref{threeone}), however, depends only on {\it one}
function of $x$ for each of the background fields since its dependence
on the background fields is inherited from the action in
(\ref{twoone}) which has this property
\footnote{The situation becomes worse when higher order corrections in
$\alpha'$ are included in the differential equations in
(\ref{twothreea})-(\ref{twothreec}) since then the order in $\eta$
derivatives in these equations increases with a corresponding increase
in the number of independent functions of $x$, for each background
field, characterizing the general solution.}.

One might worry that the functional intergral in (\ref{threeone})
defining the partition function for the noncritical string is only
formal and that a more rigorous definition of this functional integral
would show some subtility in counting the number of independent
couplings. That in fact there is no proliferation of couplings in a
more rigorous setting can be easily seen by reformulating
(\ref{threeone}) in the framework of dynamically triangulated random
surfaces. \footnote{See, for example, \cite{KM}.} Thus the
partition function in (\ref{threeone}) cannot be identified with a
(D+1)-dimensional gravitational action evaluated on-shell for general
solutions.

Actually, there are solutions to the differential equations
(\ref{twothreea})-(\ref{twothreec}) which depend on only one function
of $x$ for each background field. These solutions involve
(D+1)-dimensional spaces with a D-dimensional boundary \footnote{We
will assume the D-dimensional boundary metric to have a Euclidean
signature}, and a regularity condition in the bulk generally picks up
only one of the two possible (at the lowest order in $\alpha'$)
solutions which evolves the boundary data into the bulk. The
(D+1)-dimensional action evaluated on-shell for such solutions depends
on only {\it one} function of $x$ for each of the background fields,
which together constitute the boundary data.

A simple example of the above is provided by the coupling
corresponding to the tachyon field in the noncritical string theory
corresponding to the flat space linear dilaton solution of the
equations (\ref{twothreea})-(\ref{twothreec}). The gravitationally
dressed tachyon coupling satisfies the following equation in this
background \cite{DDW,DNW}:
\beq 
(\del^2_\eta - Q\del_\eta + \del^2_x) T(x, \eta) -
{\del V \over \del T} = 0,
\label{threeonea}
\eeq
where $V = - T^2 + O(T^3)$ is the tachyon potential. Ignoring the
cubic and higher order terms in the potential, and going to the momentum
space conjugate to $x$, we get the two solutions
\beq
{\tilde T}_\pm (k, \eta) = e^{{Q \over 2} \eta} \psi_{\pm, k}(\eta)
\label{threeoneb} 
\eeq
where $\psi_{\pm, k}(\eta)$ is the liouville wavefunction,
\beq
\psi_{\pm, k}(\eta) = e^{\pm \eta {\sqrt{(k^2 - {D-1 \over 12})}}} 
\ \psi_{\pm, k}.
\label{threeonec}
\eeq
If we place a D-dimensional ``boundary'' in the liouville direction at
$\eta_0$, we see that only the solution $\psi_-(\eta)$ is regular at
$\eta \rightarrow +\infty$. The other solution, $\psi_+(\eta)$, is
regular at $\eta \rightarrow -\infty$. The liouville wavefunctions for
both these solutions are not normalizable. \footnote{For $D \geq 1$
this is true only if $k^2 \geq {D-1 \over 12}$, and then these
wavefunctions correspond to microscopic states. Wavefunctions
corresponding to operators with $k^2 < {D-1 \over 12}$ are
normalizable and hence these correspond to macroscopic states. See,
for example, \cite{NS} for a review of liouville theory.}
 
A more nontrivial example is provided by AdS gravity, which has been
extensively studied recently in the context of the AdS/CFT
correspondence. In this case the Fefferman-Graham theorem \cite{FG}
guarantees that there is a unique regular solution that evolves
boundary data into the AdS bulk. The regular solution is, however, not
normalizable \cite{W,GKP}, just like the liouville wavefunctions for
microscopic states in the above example of tachyon in flat space and
linear dilaton background. For example, the regular solution to
linearized equations for a scalar field in the AdS background behaves
at large $r$ as $\lambda(x) r^{\Delta - D},$ where $\Delta$ is the
dimension of the corresponding operator in the dual CFT description
and $D$ is the dimension of the boundary. \footnote{The other
solution, which behaves at large $r$ as $r^{-\Delta}$, is
normalizable.}

We propose to identify the partition function in (\ref{threeone}) with
a ``boundary'' action evaluated on regular solutions obtained as described
above. The boundary in $\eta$ is the D-dimensional space parametrized
by the $x^\mu$'s and the background fields appearing in the
$\sigma$-model action $S$, (\ref{twoone}), essentially account for the
boundary values of the corresponding dressed fields. \footnote{It is
important to note here that the $\sigma$-model couplings are {\it not}
equal to the values of the dressed couplings at the boundary. The
former can, however, be traded for the latter once the dressings are
known. The example of the dressing of the tachyon in noncritical
string theory considered above provides a good illustration of
this. The tachyon coupling that enters the $\sigma$-model is $\psi_k $
which appears on the right hand side of (\ref{threeonec}). This can
clearly be traded for $\psi_k(\eta_0)$ using (\ref{threeonec}).}
Moreover, since no independent functions of $x$ corresponding to the
components $G_{\eta \eta}$ and $G_{\eta \mu}$ of the (D+1)-dimensional
metric appear in $S$, we propose that the partition function actually
evaluates the boundary action in the gauge (\ref{twofour}). As
evidence for this we point out that, by construction, the partition
function has D-dimensional general covariance, as is required by the
second of the conditions in (\ref{twofour}). A more non-trivial check
for our proposal is provided by the first gauge-fixing condition which
requires the partition function to satisfy a Hamilton-Jacobi type of
constraint equation.  Later in this section we will derive a flow
equation for the partition function which has a remarkable resemblance
to such an equation.

\vspace{2 ex}
\noindent {\bf The Liouville Boundary}
\vspace{1 ex}

Let us first try to understand how a boundary arises in the liouville
direction. As is well-known, in critical string theory, where the 2-d
metric is non-dynamical, the background fields appearing in the
$\sigma$-model action must satisfy the beta function equations for
conformal invariance. In the noncritical formulation of string theory,
however, the D-dimensional background fields appearing in
(\ref{twoone}) are completely arbitrary, since it is the integration
over the liouville mode that now enforces conformal invariance. To
make the discussion more general, let us rewrite (\ref{threeone}) as
follows:
\beq 
Z[\lambda] = \int [{\cal D} g_{\alpha \beta}] \ 
Z_g[\lambda],
\label{threetwo}
\eeq
where
\beq
Z_g[\lambda] = \int [{\cal D} X^\mu] \  
exp\bigg(\sum_i \int d^2\xi \sqrt g \ \lambda^i(X(\xi)) \  O_i(X(\xi), 
g_{\alpha \beta}(\xi))\bigg) 
\label{threethree} 
\eeq 
Here the set $\{O_i\}$ forms a complete basis of closed string
operators and so the set of couplings $\{\lambda^i\}$ includes all the
closed string modes. Now, from the point of view of the D-dimensional
matter functional integral in (\ref{threethree}), the 2-d metric is
just an external fiducial metric. As is well-known, in this case for
generic couplings $\lambda^i$, $Z_g[\lambda]$ in (\ref{threethree})
has a conformal anomaly. This anomaly has the effect that in the
generic case all the couplings $\lambda^i$ get dressed by 2-d
gravity. In perturbation theory, the liouville wavefunctions which
give the dressings of the couplings are of two types, microscopic and
macroscopic \cite{NS}, depending on the operator they couple to. Since
the wavefunctions which dress microscopic operators are not
normalizable, integration over the liouville zero mode of correlation
functions involving a sufficient number of these operators is
generically divergent. This is a source of divergent contributions to
the partition function, which thus needs a regulator to make it
well-defined.

The dressing of the coupling corresponding to the dilaton operator
plays a somewhat special role since this determines the effective
string coupling. This can bring in other problems, so let us discuss
the dressing of the dilaton coupling in some detail.

Let us first separate out the integral over the zero mode of $x$ in
(\ref{threethree}) and write
\beq 
Z_g[\lambda] = \int d^Dx \ {\cal Z}_g[\lambda; x].
\label{threefour}
\eeq 
Now, let us consider the dilaton coupling. The dressing of this
coupling is controlled by the corresponding beta function, which is
essentially the matter central charge, and is determined by the
condition that the total central charge of the matter plus 2-d gravity
system should vanish.  As a result when we fix the conformal gauge and
change over from the liouville mode to the variable $\eta(\xi)$ in the
functional integral of ${{\cal Z}}_g[\lambda; x]$
over the 2-d metric, we pick up a linear term in the action for
$\eta(\xi)$, similar to that in (\ref{twotwo}), but now with the
coefficient $Q$ given by \cite{CFMP}
\beq
Q &=& \bigg[{25-D \over 3} + \alpha' ( R^{(D)} + 
4 \nabla_\mu \nabla^\mu \Phi 
- 4 \nabla_\mu \Phi \nabla^\mu \Phi - {1 \over 12} H_{\mu \nu \lambda} 
H^{\mu \nu \lambda}) \nonumber \\ 
&&~~~~~~~~~~~~~~~~~~~~~~~~~~~~~~~~~~~~~ + O({\alpha'}^2) \bigg]^{1 \over 2}.
\label{threefive}
\eeq 
For flat space we recover the result in (\ref{twotwo}). In general
there is a linear term in $\eta(\xi)$ even in critical dimensions.
Also, since the other beta functions do not vanish for generic
couplings, $Q$ is in general a function of $x$.  The generic case is,
therefore, difficult to deal with. However, since the couplings are
arbitrary, we can choose them to be such that $Q$ is a real constant,
\footnote{$Q$ real is needed to ensure a space-like interpretation for
$\eta(\xi)$.}  independent of $x$, or at least sufficiently slowly
varying with $x$ so that its dependence on $x$ can be ignored to the
first approximation. Assuming this simplifying choice, we then have an
effective string coupling that grows at one end of the $\eta$
direction. This is true even in critical dimensions. String
perturbation theory, therefore, breaks down because of this
strong-coupling singularity.

In some cases this strong-coupling singularity can be removed by
generating a potential for $\eta(\xi)$ by switching on some additional
backgrounds. For example, in the case of flat space in noncritical
dimensions, a potential for $\eta(\xi)$ is generated by a 2-d
cosmological constant term. We will assume here that we are dealing
with such a case and an appropriate coupling has been switched on to
remove the strong coupling singularity. However, even in cases where
the strong-coupling singularity can be removed in this way,
generically there is a divergence in the partition function
(\ref{threetwo}) which comes from integration over the opposite end of
the $\eta$ direction where the effective string coupling becomes
arbitrarily weak. As we have already remarked, this is because of the
contributions to the $\sigma$-model partition function of correlators
involving microscopic loop operators whose liouville wavefunctions are
not normalizable. Thus the partition function in (\ref{threethree})
diverges for generic couplings, $\{\lambda^i\}$. \footnote{This
happens even when $Q$ in (\ref{threefive}) vanishes.} From the
world-sheet point of view this divergence is ultraviolet in
nature because it comes from 2-d surfaces of small area. However, from
the (D+1)-dimensional point of view, this is an {\it infrared}
divergence since it arises from the infinite volume in the
(noncompact) liouville direction. One way of regulating this
divergence is by introducing an appropriate cut-off on the integration
over the zero mode of $\eta(\xi)$. This is how a ``boundary'' gets
introduced in the $\eta$ direction, its location being at the value of
the cut-off.

Once a boundary has been introduced in this way, we can trade-off the
couplings appearing in (\ref{threetwo}) for the boundary values of the
dressed couplings. Now, as discussed in the previous section, a shift
in the liouville mode generates a {\it local} scale transformation on
the boundary. Therefore, a shift of the cut-off generates an RG flow
in the regularized partition function through the boundary values of
the dressed couplings, leading to a flow equation which we will now
derive.

\vspace{2 ex}
\noindent {\bf The Equation}
\vspace{1 ex}

After fixing the conformal gauge and transforming from the liouville
mode to the variable $\eta(\xi)$, the partition function in
(\ref{threetwo}) may be written as
\beq
Z[\lambda; \eta_0] = \int d^Dx \ {\cal Z}[\lambda; \eta_0; x]
\label{threeseven}
\eeq
where
\beq
{\cal Z}[\lambda; \eta_0; x]
= \int^{\infty}_{\eta_0} d\eta \ {\cal L} [\lambda(\eta); x],
\label{threeeight}
\eeq
and $\eta$ is the zero mode of $\eta(\xi)$ with $\eta_0$ the cut-off 
or the boundary value. On the right hand side of (\ref{threeeight})
we have made it explicit that the $\eta$-dependence comes entirely from  
the dressings of the couplings.

The flow equation can now be derived by making an $x$-dependent change
in $\eta_0$, namely $\eta_0 \rightarrow \eta_0 + \epsilon(x)$. 
\footnote{This $x$-dependent change in $\eta_0$ is made possible by
the fact that we could have chosen an $x$-dependent cut-off on
$\eta$. This possibility of a {\it local} cut-off on $\eta$ is
compatible with the requirement of world-sheet reparametrization
invariance.} Denoting the value of the dressed coupling at the
boundary by $\lambda^i_0(x)$, the flow equation is
\beq 
{\cal L} [\lambda(\eta_0); x] = 
\del_{\eta_0}\lambda^i_0(x) {\delta Z \over
\delta \lambda^i_0(x)} 
\label{threenine}
\eeq 
This equation follows from the fact that a change in ${\cal Z}
[\lambda; \eta_0; x] $ produced by a shift in $\eta_0$ can be computed
in two different ways. One is directly from the way $\eta_0$ appears
as an integration limit in (\ref{threeeight}). This gives the left
hand side of the equation.  The other is by recognizing that ${\cal Z}
[\lambda; \eta_0; x]$ depends on $\eta_0$ only through the boundary
values of the dressed couplings since the $\eta$ dependence in ${\cal
L} [\lambda(\eta); x]$ comes entirely from the liouville dressings of
the couplings and the subsequent transformation to the gaussian
variable $\eta$. This gives the right hand side of the equation.

At the lowest order in $\alpha'$, we expect ${\cal L} [\lambda(\eta);
x]$ to be only quadratic in $\eta$-derivatives of the dressed
couplings. Let us write this out explicitly as
\footnote{We may assume the standard normalization for the ``kinetic''
term without any loss of generality. A possible linear term in
$\lambda^i(x, \eta)$ can be removed by a field redefinition of the
original sigma-model couplings ${\lambda^i}$. We will assume that this
has been done and that the couplings ${\lambda^i}$ have been chosen
accordingly. Also, note that the right hand side of (\ref{threeten})
is evaluated on-shell in the sense that the $\eta$-dressing of the
various couplings is determined by the requirement of
reparametrization invariance. We also mention that the form of ${\cal
L}$ assumed in (\ref{threeten}) can be derived close to a CFT point in
theory space.}
\beq
{\cal L} [\lambda(\eta); x] = {1 \over 2} {\cal
G}_{ij} \del_{\eta} \lambda^i(x, \eta) \del_{\eta} \lambda^j(x, \eta)
+ V[\lambda(\eta); x]
\label{threeten}
\eeq
where $\lambda^i(x, \eta)$ is the dressed coupling, ${\cal G}^{ij}$ is

the metric on the space of the couplings and $V$ is assumed to have a
local expansion in $x$-derivatives of $\lambda^i(x, \eta)$. We have
made the reasonable assumption that ${\cal L}$ has a low energy
expansion in derivatives of $\eta$ and $x$. Once (\ref{threeten}) is
given, one can show that the variation of the partition function with
respect to the boundary values of the dressed couplings is related to
the ``velocities'' in the standard way,
\beq
{\delta Z \over \delta \lambda^i_0(x)} = 
{\cal G}_{ij} \del_{\eta_0} \lambda^j_0(x).
\label{threeeleven}
\eeq
The flow equation (\ref{threenine}) may then be rewritten as
\beq
{1 \over 2} {\cal G}^{ij} {\delta Z \over \delta \lambda^i_0(x)}
{\delta Z \over \delta \lambda^j_0(x)} = V[\lambda_0; x] 
\label{threetwelve}
\eeq
which is the advertized Hamilton-Jacobi type of constraint equation that the
regularized partition function must satisfy.

\section{CONCLUDING REMARKS}

In this note we have presented noncritical string theory as a boundary
value problem, based on the observation that the liouville or
conformal mode gives rise to an additional dimension. As we have
argued, the boundary arises from the cut-off needed to regulate
world-sheet ultraviolet divergences. We have shown that, under some
reasonable assumptions, the partition function of the noncritical
string $\sigma$-model action satisfies a Hamilton-Jacobi type of
constraint equation as a functional of the boundary values of the
$\sigma$-model couplings. The dependence of the couplings on the
additional dimension is determined by the first order {\it local} RG
flow equations (\ref{threeeleven}). These equations were obtained for
the bosonic string, but extension to the superstring is
straightforward when RR backgrounds are absent.  Since RR backgrounds
couple to bilinears of space-time fermions in the $\sigma$-model, the
analysis becomes complicated when these backgrounds are switched on
\cite{BVW}. For this reason it is difficult to demonstrate explicitly
that a Hamilton-Jacobi type of constraint equation continues to be
satisfied in the presence of RR backgrounds, although we expect this
to be the case. Finally we mention that the structure of the solution
space of the RG flow equations (\ref{threeeleven}) is presently not
known \cite{Gub}. In order to address this issue it would be
worthwhile to discuss the global topology of the RG flows along the
lines presented in \cite{DMW} where the global topology of a class of
$c<1$ models was exactly calculated using methods of Morse theory.

\noindent {\bf Acknowledgements}

\noindent One of us (SRW) would like to thank Theory Division, CERN, for
hospitality during a visit when part of this work was done.

\end{document}